\documentstyle[12pt,epsfig]{article}
\begin{document}

\begin{center}
{\Large\bf
(Anti-)self-dual homogeneous gluon field and axial anomaly in QCD}
\vspace*{.5cm}

{\bf Ja. V. Burdanov}, \\
Laboratory of Nuclear Problems,\\
Joint Institute for Nuclear Research, Dubna, Russia \\
{\bf G. V. Efimov and S. N. Nedelko}, \\
Bogoliubov Laboratory of Theoretical Physics, \\
Joint Institute for Nuclear Research, Dubna, Russia \\

\end{center}

\begin{abstract}
        The transition form factor $F_{\gamma\pi}(Q^2)$,
        decay width $\Gamma(\pi^0\to\gamma\gamma)$ and
        charge form factor of pion $F_{\pi}(Q^2)$ are calculated
        within the model of induced nonlocal quark currents
        based on the assumption that the nonperturbative QCD vacuum
        can be characterized by a homogeneous (anti-)self-dual gluon field.
        It is shown that the interaction of the quark spin with the
        vacuum gluon field, being responsible for the chiral symmetry
        breaking and the spectrum of light mesons, can also play the
        decisive role in forming the form factor $F_{\gamma\pi}(Q^2)$ and
        decay width $\Gamma(\pi^0\to\gamma\gamma)$.
        An asymptotic behavior of quark loops in the presence of the
        background gluon field for large  $Q^2$ is discussed.
\end{abstract}

\section{Electromagnetic interactions within
         the model \ \ \ \  of induced nonlocal quark currents}

The model of induced nonlocal quark currents~\cite{efned,bur}
is based on the assumption that the (anti-)self-dual homogeneous
field~\cite{leutw}
\begin{eqnarray}
\label{field}
\hat B_\mu(x)=\hat n B_{\mu\nu}x_\nu, \
\hat n=\lambda_3\cos\xi + \lambda_8\sin\xi,
\nonumber\\
\tilde B_{\mu\nu}=\pm B_{\mu\nu}, \
B_{\mu\rho}B_{\rho\mu}=-B^2\delta_{\mu\nu}
\end{eqnarray}
can be considered as a dominating gluon configuration in the QCD vacuum.

Interaction of quarks with electromagnetic field $A_\mu$
can be introduced within the model by means of the minimal
substitution (see also~\cite{gross})
both in the free and interaction parts of quark Lagrangian.

By means of the bosonization procedure~\cite{efned,bur}
the functional integral $Z[A]$ in the presence of electormagnetic field
$A_\mu$ can be represented in terms of composite meson fields
$\Phi_{\cal N}$~\cite{hep}
\begin{eqnarray}
\label{genlast}
&&Z[A]=N\int \prod_{\cal N}D\Phi_{\cal N}
\exp\left\{\frac{1}{2}\int\!\!\int {\rm d}^4x{\rm d}^4y
\Phi_{\cal N}(x)\left[\left(
\Box-M_{\cal N}^2\right)\delta(x-y)\right.\right. \nonumber\\
&&-\left.\left.h_{\cal N}^2\Pi_{\cal N}^R(x-y)\right]
\Phi_{\cal N}(y)+I_{\rm int}\left[\Phi\mid A\right]\right\}, \\
&&I_{\rm int}=
-\int d^4x h_{\cal N}
\Phi_{\cal N}(x)\left[
\Gamma_{\cal N}(x\mid A)-
\Gamma_{\cal N}(x\mid 0)
\right]
\nonumber\\
&&-\frac{1}{2}\int d^4x_1\int d^4x_2 h_{\cal N}h_{\cal N^\prime}
\Phi_{\cal N}(x_1)\left[\Gamma_{\cal N N'}(x_1,x_2\mid A)-
\delta_{\cal NN'}\Pi_{\cal N}(x_1-x_2)\right]\Phi_{\cal N'}(x_2)
\nonumber\\
&&-\sum_{m=3}\frac{1}{m}\int d^4x_1...\int d^4x_m
\prod_{k=1}^mh_{{\cal N}_k}\Phi_{{\cal N}_k}(x_k)
\Gamma_{{\cal N}_1...{\cal N}_m}(x_1,...,x_m\mid A),\nonumber\\
\label{Gamma}
&&\Gamma_{{\cal N}_1...{\cal N}_m}=\int d\sigma_{\rm B}
{\rm Tr} V_{{\cal N}_1}(x_1\mid A)S(x_1,x_2\mid A)...
V_{{\cal N}_m}(x_m\mid A)S(x_m,x_1\mid A),
\end{eqnarray}
where the vertex functionals $\Gamma_{{\cal N}_1...{\cal N}_m}$
include electromagnetic field $A$ through the covariant derivatives
\begin{eqnarray}
\label{vertex3}
&&V^{bJ\ell n}(x\mid A)=
M^b\Gamma^J\left\{\!\!\left\{F_{n\ell}
\left(\frac{\stackrel{\leftrightarrow}{D}^2(x)}{\Lambda^2}\right)
T^{(\ell)}\left(\frac{1}{i}
\frac{\stackrel{\leftrightarrow}{D}(x)}{\Lambda}\right)
\right\}\!\!\right\}, \\
\label{D}
&&\stackrel{\leftrightarrow}{D}^\mu_{ff^\prime}=
\xi_f\left[\stackrel{\leftarrow}{\nabla}^\mu+ie_fA^\mu(x)\right]-
\xi_{f^\prime}\left[\stackrel{\rightarrow}{\nabla}^\mu
-ie_{f^\prime}A^\mu(x)\right].
\nonumber
\end{eqnarray}
The action in~(\ref{genlast}) is invariant under $U(1)$ transformations.

Meson-quark coupling constants are defined by the relations:
\begin{eqnarray}
\label{h}
h^2_{\cal N}=
1/\tilde\Pi^\prime_{\cal N}(p^2)|_{p^2=-M^2_{\cal N}}.
\end{eqnarray}
To get various meson-photon amplitudes one has to decompose the vertex
functionals $\Gamma_{{\cal N}_1...{\cal N}_m}$
into a series over the electromagnetic field $A$.

Fourier transform of the translation invariant part $H_f$
of the quark propagator in the homogeneous (anti-)self-dual gluon field
reads~\cite{bur,eur}
\begin{eqnarray}
\label{qsol}
\tilde H_f(p)=
\frac{1}{2v\Lambda^2}\int\limits_{0}^{1}ds
e^{-\frac{p^2}{2v\Lambda^2}s}\left(\frac{1-s}{1+s}\right)
^{\frac{m_f^2}{4v\Lambda^2}}
\left[
p_\alpha\gamma_\alpha\pm is\gamma_5\gamma_\alpha f_{\alpha\beta}p_\beta
\right.
\nonumber \\
\left.
+m_f\left(P_\pm +P_\mp
\frac{1+s^2}{1-s^2}-\frac{i}{2}\gamma_\alpha f_{\alpha\beta}\gamma_\beta
\frac{s}{1-s^2}\right)\right].
\end{eqnarray}
Contribution of zero modes to the propagator
is seen in Eq.~(\ref{qsol}) as a singilarity of the integrand
at $s=1$ which is integrable unless $m_f=0$.
Zero modes are due to interaction of a quark spin with the vacuum field.

\section{Decay $\pi^0\to \gamma\gamma$
and $\gamma^{\ast}\pi^0\to\gamma$ transition form factor}

The contribution of the triangle diagram to the form factor
$F_{\gamma\pi}(Q^2)$ can be represented as an integral over proper times
$t$ and $s_1$, $s_2$, $s_3$, corresponding to the vertex and
propagators, respectively~\cite{hep}:
\begin{eqnarray}
\label{5}
F_{\gamma\pi}(Q^2)&=&\frac{1}{\Lambda}\frac{h_\pi}{2\sqrt{2}\pi^2}
{\rm Tr}_v\int \limits_{0}^{1}...\int \limits_{0}^{1}dtds_1ds_2ds_3
\left[\left(\frac{1-s_1}{1+s_1}\right)
\left(\frac{1-s_2}{1+s_2}\right)
\left(\frac{1-s_3}{1+s_3}\right)\right]^{\frac{m^2_u}{4v}}
\nonumber\\
&\times&
\sum\limits_{i=1,2,3}\frac{m_u}{1-s_i^2}\Phi_i(M^2_\pi,Q^2;s,t)
\exp\left[M^2_\pi\phi(s,t)-Q^2\varphi(s,t)\right].
\end{eqnarray}
The singularities $(1-s_i)^{-1}$ of the integrand in Eq.~(\ref{5})
at $s_i\to 1$ appear from the zero mode contribution
to the quark propagator (see the second line in Eq.~(\ref{qsol})).
These singularities lead to the $1/m_u$-dependence of the integral
in Eq.~(\ref{5}) in the limit $m_u\ll \Lambda$, while the effective
coupling constant $h_\pi$ in Eq.~(\ref{h}) behaves as $h_\pi\sim m_u$,
and $F_{\gamma\pi}(Q^2)$ does not vanish as $m_u\to 0$ but approaches
a constant value.
The same is valid for decay width $\Gamma(\pi^0\to\gamma\gamma)$.
This asymptotic analysis illustrates an appearance of axial anomaly
as a result of interaction of a quark spin with the vacuum field.

Numerical results for the form factor and decay width
are represented in Fig.~a and Table~1.
The model parameters are fitted from the meson masses:
$m_u=m_d=198.3 \ {\rm MeV}, \Lambda=319.5 \ {\rm MeV}$ that gives
$M_\pi=140 \ {\rm MeV}, h_\pi=6.51$~\cite{bur}.
The radius for $\gamma^{\ast}\pi^0\to\gamma$ transition
is equal to $.57~{\rm fm}$, that have to be compared with
$r_{\gamma\pi}^{\rm exp}=.65\pm.03~{\rm fm}$~\cite{bebek}.

Within the model of induced nonlocal currents the
$\gamma^{\ast}\pi^0\to\gamma$ transition form factor and two photon
decay constant, a smallness of pion mass,
splitting of the masses of vector and pseudoscalar light mesons and
weak decay constants of pions and kaons~\cite{bur}
are explained by the same reason -- an interaction of a quark spin with
the vacuum homogeneous gluon field.

A behaviour of the triangle diagram for the transition form factor
in the limit $Q^2\gg\Lambda^2$ can be easily estimated,
using the Laplace method. As a result, we arrive at relation
\begin{eqnarray}
\label{10}
&&Q^2F_{\gamma\pi}(Q^2/\Lambda^2)=
C\Lambda + O(\Lambda^2/Q^2)
\approx 0.2 \ {\rm Gev} + O(\Lambda^2/Q^2),
\end{eqnarray}
that is consistent with the analysis within factorization hypothesis
and QCD sum rule approaches~\cite{rad1}.
This result has to be compared with the Brodsky-Lepage limit~\cite{br}
$Q^2F_{\gamma\pi}(Q^2) \to 2F_\pi = .186 \ {\rm Gev}.$

\section{The pion charge form factor}

According to the effective action~(\ref{Gamma}),
the one-loop amplitude for the
processes $\pi^\pm\gamma^\ast\to\pi^\pm$
is described by the triangle and bubble diagrams.
The analitical expressions for the contributions of these diagrams
to the pion charge form factor can be found in~\cite{hep}.

For the parameter values given in~\cite{bur}
the charge form factor is plotted in Fig.~b by the solid line.
The electromagnetic radius takes the value $r_{\pi}=.524~{\rm fm}$
that is in agreement with experimental data $r_\pi^{\rm exp}=.656~{\rm fm}$.
Numerically the contribution of the bubble diagrams is
negligible at small and intermediate $Q^2$, but is the leading
contribution for the asymptotically large $Q^2$.

In our case, a naive estimation of the asymptotic behaviour of the
triangle diagram based on the ultraviolet behaviour
of the quark propagator~(\ref{qsol}) and vertex (\ref{vertex3})
gives $(Q^2)^{-2}$.

The asymptotic formula for the triangle diagram in the limit
$Q^2\gg \Lambda^2$ for the parameter values from~\cite{bur} reads
\begin{eqnarray}
\label{as}
&&F_\pi^\triangle(Q^2/\Lambda^2)=
\frac{2.96}{\left(Q^2/\Lambda^2\right)^{1.1435}}.
\end{eqnarray}
This is clearly due to the specific interplay of
translation and color gauge invariance in the quark loops
in the presence of vacuum homogeneous gluon field.
Asymptotic formula~(\ref{as}) fits well the solid curve in Fig.~b
for $Q^2> 5 \ {\rm Gev}^2$.

The contribution of bubble graph to asymptotic behaviour takes the form
\begin{eqnarray}
\label{asb}
&&F_\pi^\circ(Q^2/\Lambda^2)\approx 0.3
\Lambda^2/Q^2+O\left(\left(\Lambda^2/Q^2\right)^2\right).
\nonumber
\end{eqnarray}
The asymptotic behaviour of the pion form factor is
determined by the one-gluon exchange between quarks inside a pion.
This is consistent with the mechanism of hard rescattering and quark
counting rules~\cite{matv}.
However, in the experimentally observed region the triangle diagram
dominates in the form factor.

\newpage

\begin{table}
\begin{center}
\begin{minipage}{120.mm}
\caption{The two-photon decay constant $g_{\pi\gamma\gamma}$
(${\rm Gev}^{-1}$) and decay width $\Gamma(\pi^0\to\gamma\gamma)$
($\rm ev$); $g_{\pi\gamma\gamma}^\ast, \Gamma^\ast$ are the values
calculated without taking into account the spin-field interaction}
\end{minipage}

~\\[1cm]
\begin{tabular}{|c|c|c||c|c|c|}\hline
$g_{\pi\gamma\gamma}$&
$g_{\pi\gamma\gamma}^\ast$ &$g_{\pi\gamma\gamma}^{\rm exp}$ \cite{bebek}&
$\Gamma$&$\Gamma^\ast$&$\Gamma^{\rm exp}$ \cite{bebek}
\\ \hline
0.235&0.108&0.276&6.3&1.34&8.74
\\ \hline
\end{tabular}
\end{center}
\end{table}

\setcounter{figure}{0}
\unitlength=0.2pt
     \begin{figure}
~\\[8.cm]
     \begin{picture}(0,10)
     \put(0,10){\epsfig{file=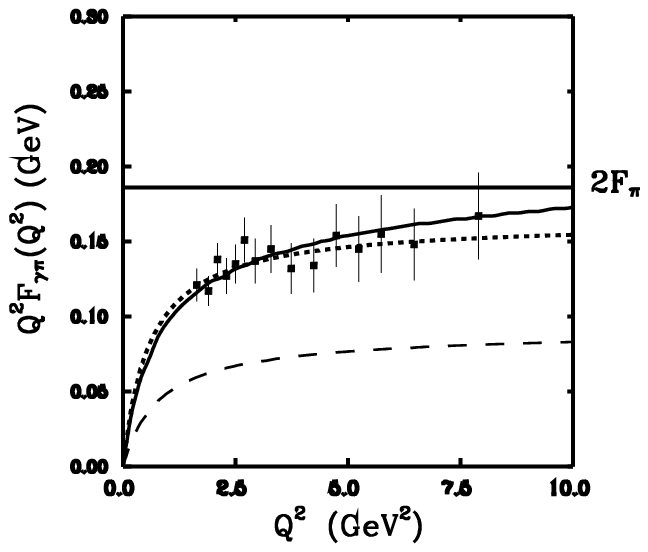,width=7.2cm}}
     \put(1200,10){\epsfig{file=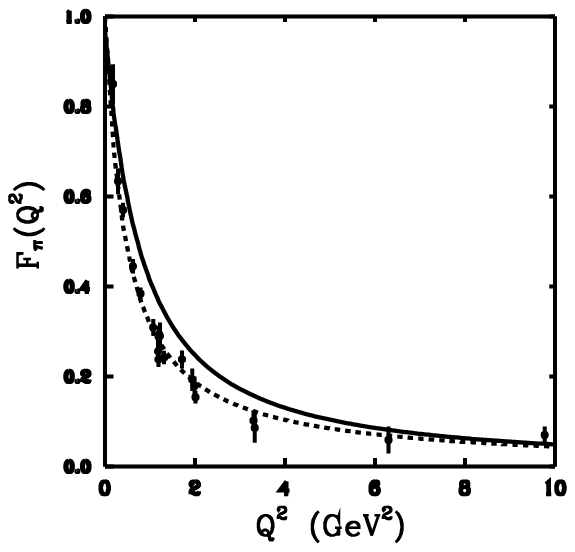,width=6.4cm}}
     \end{picture}        
~\\[-7.cm]
     \hspace*{3.5cm}(a)\hspace*{7.7cm}(b)
     \end{figure}        
~\\[6.cm]
     Transition pion form factor (a).
     Solid curve represents a result of
     the model of induced nonlocal quark currents.
     Long dashed line - calculation without taking into account
     the spin-field interaction.
     Experimental fit\protect\cite{bebek} is given by short dashed line,
     and Brodsky-Lepage limit\protect\cite{br} ($\approx .186 \ {\rm Gev}$)
     is shown by solid straight line

~\\[.5cm]
     The pion charge form factor (b) calculated
     in the present model (solid line) compared with experimental fit
     (dashed line)

\end{document}